\documentstyle[12pt,psfig,epsfig]{article}

\newcommand{\gsim}{\raisebox{-0.13cm}{~\shortstack{$>$ \\[-0.07cm] $\sim$}}~}

\newcommand{\ra}{\rightarrow}
\newcommand{\ee}{e^+e^-}
\newcommand{\s}{\\ \vspace*{-3mm} }
\newcommand{\nn}{\noindent}
\newcommand{\non}{\nonumber}
\newcommand{\beq}{\begin{eqnarray}}
\newcommand{\eeq}{\end{eqnarray}}
\newcommand{\SM}{\mbox{${\cal SM}$}}

\newcommand{\MSSM}{\mbox{${\cal MSSM}$}}

\newcommand{\tb}{\mbox{tg$\beta$}}

\begin{document}

\begin{titlepage}

\begin{flushright}
DESY 95--214\\
KA--TP--10--95\\ 
SCIPP 95/47\\
%December 1995 \\
\end{flushright}

\vspace{1cm}

\def\thefootnote{\fnsymbol{footnote}}

\begin{center}

{\large\sc {\bf Multiple Production of \MSSM\ Neutral Higgs Bosons}} 

\vspace*{3mm}

{\large\sc {\bf at High--Energy e$^+$e$^-$ Colliders}}

\vspace{1cm}

{\sc A.~Djouadi$^{1,2}$\footnote{Supported by Deutsche Forschungsgemeinschaft 
DFG (Bonn).}, H.E.~Haber$^3$, and P.M.~Zerwas$^2$ } 

\vspace{1cm}

$^1$ Institute f\"ur Theoretische Physik, Universit\"at Karlsruhe, \\
D--76128 Karlsruhe, FRG. \\
\vspace{0.3cm}

$^2$ Deutsches Elektronen--Synchrotron DESY, D-22603 Hamburg, FRG. \\
\vspace{0.3cm}

$^3$ Santa Cruz Institute for Particle Physics, University of California, \\
Santa Cruz, CA 95064, USA. \\
\vspace{0.3cm}

\end{center}

\vspace{1.6cm}

\begin{abstract}
\normalsize
\noindent

\nn The cross sections for the multiple production of the lightest
neutral Higgs boson at high--energy $e^+ e^-$ colliders are presented in
the framework of the Minimal Supersymmetric extension of the Standard
Model (\MSSM). We consider production through Higgs--strahlung,
associated production of the scalar and the pseudoscalar bosons, and the
fusion mechanisms for which we use the effective longitudinal
vector--boson approximation. These cross sections allow one to determine
trilinear Higgs couplings $\lambda_{Hhh}$ and $\lambda_{hhh}$, which are
theoretically determined by the Higgs potential. 

\end{abstract}

\end{titlepage}

\def\thefootnote{\arabic{footnote}}
\setcounter{footnote}{0}
\setcounter{page}{1}

\subsection*{1. Introduction}

The only unknown parameter in the Standard Model (\SM) is the quartic coupling
of the Higgs field in the potential, which determines the value of the Higgs
mass. If the Higgs mass is known, the potential is uniquely fixed. Since the
form of the Higgs potential is crucial for the mechanism of spontaneous
symmetry breaking, {\it i.e.} for the Higgs mechanism {\it per se}, it will be
very important to measure the coefficients in the potential once Higgs
particles have been discovered. \s 

If the mass of the scalar particle is less than about 150 GeV, it very
likely belongs to the quintet of Higgs bosons,
$h,H,A,H^\pm$ predicted in the
two--doublet Higgs sector of supersymmetric theories \cite{R1}
[$h$ and $H$ are the light and heavy CP--even Higgs bosons, $A$ is the CP--odd 
(pseudoscalar) Higgs boson, and $H^\pm$ is the charged Higgs pair]. The
potential of the two doublet Higgs fields, even in the Minimal Supersymmetric
Standard Model (\MSSM), is much more involved than in the Standard Model
\cite{X2}. If CP is conserved by the potential, the most general
two--doublet model contains three mass parameters and seven real
self--couplings. In the \MSSM, the potential automatically conserves CP;
in addition, supersymmetry fixes all the Higgs
self--couplings in terms of gauge couplings.  The remaining three
free mass parameters can be traded in for the two vacuum expectation 
values (VEV's) of the neutral Higgs fields and one of the physical Higgs masses.
The sum of the squares of the VEV's is fixed by the $W$ mass, while
the ratio of VEV's is a free parameter of the model called $\tb$.  It is
theoretically convenient to choose the free parameters of the \MSSM\ Higgs
sector to be $\tb$ and $M_A$, the mass of the CP--odd Higgs boson $A$.
The other Higgs masses and the mixing angle $\alpha$ of the CP--even neutral
sector are then determined. Moreover, since all coefficients in the
Higgs potential are also determined, the trilinear and
quartic self--couplings of the physical Higgs particles can be
predicted theoretically. By measuring these couplings, the Higgs potential can
be reconstructed -- an experimental {\it prima  facie} task to establish the
Higgs mechanism as the basic mechanism for generating the masses of the
fundamental particles. \s 

The endeavor of measuring all Higgs self--couplings in the \MSSM\ is a
daunting task. We will therefore discuss a first step by analyzing
theoretically the production of two light Higgs particles of the \MSSM.
These processes may be studied at the proton collider LHC \cite{X2a} and at a
high--energy $\ee$ linear collider. In this paper we will focus on the
$\ee$ accelerators that are expected to operate in the first phase at an
energy of 500 GeV with a luminosity of about $\int {\cal L} = 20$
fb$^{-1}$, and in a second phase at an energy of about 1.5 TeV with a
luminosity of order $\int {\cal L} =200$ fb$^{-1}$ {\it per annum}
\cite{X3}. They will allow us to eventually study the couplings 
$\lambda_{Hhh}$ and $\lambda_{hhh}$. The measurement of the coupling 
$\lambda_{hAA}$ will be very difficult. \s 

Multiple light Higgs bosons $h$ can [in principle] be generated in the \MSSM~by
four mechanisms\footnote{The production of two light Higgs bosons, $\ee \ra
hh$, through loop diagrams does not involve any trilinear Higgs 
coupling; the production rates are rather small \cite{R3}.}: \s

\noindent (i) \underline{Decay of the heavy CP--even neutral Higgs boson}, 
produced either by $H$--strahlung and associated $AH$ pair production, 
or in the $WW$ fusion mechanisms, Fig.~1a, 
\begin{eqnarray}
\left. 
\begin{array}{l}
\ee \ra ZH, \ AH \\
\ee \ra \nu_e \bar{\nu}_eH 
\end{array}
\right\} \hspace*{.6cm} H \ra hh
\end{eqnarray}
Associated production $\ee \ra hA$ followed by $A \ra hZ$ decays leads to 
$hhZ$ background final states. \s

\nn (ii) \underline{Double Higgs--strahlung in the continuum}, with a final 
state $Z$ boson, Fig.~1b, 
\begin{eqnarray}
\ee \ra Z^* \ra hhZ
\end{eqnarray}
\nn (iii) \underline{Associated production with the pseudoscalar $A$ in the 
continuum}, Fig.~1c, 
\begin{eqnarray}
\ee \ra Z^* \ra hhA
\end{eqnarray}
(iv) \underline{Non--resonant $WW(ZZ)$ fusion in the continuum}, Fig.~1d, 
\begin{eqnarray}
\ee \ra \bar{\nu}_e \nu_e W^* W^* \ra \bar{\nu}_e \nu_e hh 
\end{eqnarray}
The cross sections for $ZZ$ fusion in (1) and (4) are suppressed by an order 
of magnitude.
The largest cross sections can be anticipated for the processes (1), where
heavy on--shell $H$ Higgs bosons decay into pairs of the light Higgs bosons.
[Cross sections of similar size are expected for the backgrounds involving the
pseudoscalar Higgs bosons.] We have derived the cross sections for the four
processes analytically; the fusion process has been treated in the equivalent
particle approximation for longitudinal vector bosons. \s 

We will carry out the analysis in the \MSSM~for the value $\tb=1.5$. [A
summary will be given in the last section for all values of $\tb$]. In
the present exploratory study, squark mixing will be neglected, {\it
i.e.} the supersymmetric Higgs mass parameter $\mu$ and the parameter
$A_t$ in the soft symmetry breaking interaction will be set to zero, and
the radiative corrections will be included in the leading $m_t^4$ one--loop
approximation parameterized by \cite{R4} 
\begin{eqnarray}
\epsilon = \frac{3G_F}{\sqrt{2} \pi^2} \frac{m_t^4}{\sin^2 \beta} 
\log	\left( 1+ \frac{M_S^2}{m_t^2} \right)
\end{eqnarray}
with the common squark  mass fixed to $M_S=1$ TeV. 
In terms of $\tb$ and $M_A$, the trilinear Higgs couplings relevant 
for our analysis are given in this approximation by 
\begin{eqnarray}
\lambda_{hhh} &=& 3 \cos2\alpha \sin (\beta+\alpha) 
+ 3 \frac{\epsilon}{M_Z^2} \frac{\cos^3 \alpha}{\sin\beta}  \\
\lambda_{Hhh} &=& 2\sin2 \alpha \sin (\beta+\alpha) -\cos 2\alpha \cos(\beta
+ \alpha) + 3 \frac{\epsilon}{M_Z^2} \frac{\sin \alpha}{\sin\beta}
\cos^2\alpha \non 
\end{eqnarray}
In addition, the coupling 
\begin{eqnarray}
\lambda_{hAA} &=& \cos 2\beta \sin(\beta + \alpha) + \frac{\epsilon}{M_Z^2} 
\frac{\cos \alpha}{\sin\beta} \cos^2\beta 
\end{eqnarray}
will be needed even though it turned out -- {\it a posteriori} -- that it
cannot be measured using the experimental methods discussed in this 
note\footnote{For small masses the decay $h \ra AA$ could have provided 
an experimental opportunity to measure this coupling. However, for
$\tb >1$, this area of the \MSSM\ parameter space has been excluded
by LEP \cite{grivaz}.}. As
usual, these couplings are defined in units of $( 2\sqrt{2} G_F)^{1/2} M_Z^2$;
the $h,H,H^\pm$ masses and the mixing angle $\alpha$ can be expressed in terms
of $M_A$ and $\tb$ [see e.g. Ref.~\cite{R5} for a recent discussion]. \s 

In the decoupling limit \cite{R8} for large $A$, $H$ and $H^\pm$ masses, the
lightest Higgs particle becomes \SM--like and the trilinear $hhh$ coupling
approaches the \SM~value $\lambda_{hhh} \ra M_h^2/M_Z^2$.
In this limit, only the first three diagrams of Fig.~1b and 1d contribute
and the cross-sections for the processes $\ee \ra hhZ$ and $WW \ra hh$
approach the corresponding cross sections of the \SM\ \cite{R6,R7}.

\subsection*{2. H Production and hh Decays}

If kinematically allowed, the most copious source of multiple $h$ 
final states are cascade decays $H \ra hh$, with $H$ produced either
by Higgs--strahlung or associated pair production \cite{R1},
\beq
\sigma (\ee \ra ZH) &=& \frac{G_F^2 M_Z^4}{96 \pi s } (v_e^2+a_e^2) 
\cos^2(\beta-\alpha) \frac{ \lambda^{1/2}_Z [\lambda_Z+12M_Z^2/s]}  
{(1-M_Z^2/s)^2}   \\
\sigma (\ee \ra AH) &=& \frac{G_F^2 M_Z^4}{96 \pi s} (v_e^2+a_e^2)
\sin^2(\beta-\alpha) \frac{ \lambda^{3/2}_A} {(1-M_Z^2/s)^2} 
\eeq
The $Z$ couplings to electrons are given by $a_e=-1, v_e=-1+ 4\sin^2\theta_W$ 
and $\lambda_j$ is the usual two--body phase space function $\lambda_{j}
= (1-M_j^2/s-M_H^2/s)^2-4M_j^2 M_H^2/s^2$. The cross sections (8) and (9) are 
shown in Fig.~2 for the total $\ee$ 
energies $\sqrt{s}=500$ GeV and 1.5 TeV as a 
function of the Higgs mass $M_H$ for a small value of $\tb=1.5$ where 
the $H$ cascade decays are significant over a large mass range. As a 
consequence of the decoupling theorem, associated $AH$ production is dominant
for large Higgs masses. \s

The trilinear $Hhh$ coupling can be measured in the decay process $H\ra hh$
\begin{eqnarray}
\Gamma(H \ra hh ) = \frac{G_F \lambda^2_{Hhh} }{16\sqrt{2} \pi} 
\frac{M_Z^4}{M_H}  \left(1-\frac{4M_h^2}{M_H^2} \right)^{1/2} 
\end{eqnarray}
if the branching ratio is neither too small nor too close to unity. This is 
indeed the case, as shown in Fig.~3a, for $H$ masses between 180 and 350 
GeV
and small to moderate $\tb$ values. The
other important decay modes are $WW^*/ZZ^*$ decays. Since the $H$
couplings to the gauge bosons can be measured through the production
cross sections of the fusion and Higgs--strahlung processes, the
branching ratio BR$(H \ra hh)$ can be exploited to measure the coupling
$\lambda_{Hhh}$. \s 

The $ZH$ final state gives rise to resonant two--Higgs $[hh]$ 
final states.  The $AH$ final state typically yields
three Higgs $h[hh]$ final states 
%[which includes a resonant two--Higgs pair]
since the channel $A \ra hZ$ is the dominant decay mode in most of 
the mass range we consider. This is shown in Fig.~3b where the 
branching ratios of the pseudoscalar $A$ are displayed for $\tb=
1.5$. \s

Another type of two--Higgs $hh$ final states is generated in the chain
$\ee \ra Ah \ra [Zh]h$, which does not involve any of the Higgs
self--couplings. However, in this case, the two $h$ bosons do not
resonate while $[Zh]$ does, so that the topology of these background
events is very different from the signal events. The size of the $\ee
\ra hA$ background cross section is shown in Fig.~2 together 
with the signal cross sections; for sufficiently large $M_A$, it becomes
small, in line with the decoupling theorem \cite{R8}. \s 

A second large signal cross section is provided by the $WW$ fusion
mechanism. [Since the NC couplings are smaller compared to the CC
couplings, the cross section for the $ZZ$ fusion processes in (1) and 
(4)  is $\sim
16\cos^4 \theta_W$, {\it i.e.} one order of magnitude smaller than for
$WW$ fusion.] In the effective longitudinal $W$ approximation 
\cite{Wlumi} one obtains 
\beq
\sigma( \ee \ra H \bar{\nu}_e \nu_e ) = \frac{G_F^3 M_W^4}{4 \sqrt{2}\pi} 
\left[ \left(1+\frac{M_H^2}{s} \right) \log \frac{s}{M_H^2} -2 \left(1-
\frac{M_H^2}{s} \right) \right] \cos^2(\beta-\alpha) 
\eeq
The magnitude of the cross section\footnote{In the effective $W$ 
approximation, the cross section may be overestimated by as much as a 
factor of 2 for small masses and/or small c.m. energies.
%This, however, is not dramatic in the present case, since the coupling
%$\lambda_{Hhh}$ will be extracted from the $H$ decay branching ratio.
Therefore we display the exact cross sections \cite{EE500} in Fig.2.}
$\ee \ra H\nu_e \bar{\nu}_e$ is also shown in Fig.~2 for the two energies
$\sqrt{s}=500$ GeV and 1.5 TeV as a function of the Higgs mass $M_H$ and
for $\tb=1.5$. The signals in $\ee \ra [hh]$ + missing energy are very
clear, competing only with $H$--strahlung and subsequent neutrino decays
of the $Z$ boson. Since the lightest Higgs boson will decay mainly into
$b\bar{b}$ pairs, the final states will predominantly include four and
six $b$ quarks. \s 

At $\sqrt{s}=500$ GeV, about 500 signal events are predicted in the
mass range of $M_H \sim 200$ GeV for an integrated luminosity of $\int
{\cal L}=20$ fb$^{-1}$ {\it per annum}; and at $\sqrt{s}=1.5$ TeV, about
8,000 to 1,000 signal events for the prospective integrated luminosity
of $\int {\cal L}=200$ fb$^{-1}$ {\it per annum} in the interesting mass
range between 180 and 350 GeV. Note that for both energies, the $Ah$ 
background cross section is significantly smaller. 

\subsection*{3. Non-Resonant Double hh Production}

The double Higgs--strahlung $\ee \ra Zhh$, the triple Higgs production
process $\ee \ra Ahh$ and the $WW$ fusion mechanism $\ee \ra \nu_e \bar{\nu}_e
hh$ outside the resonant $H \ra hh$ range are disfavored by an
additional power of the electroweak coupling compared to the resonance
processes. Nevertheless, these processes must be analyzed carefully in order to
measure the value of the $hhh$ coupling. 
\subsubsection*{3.1 $e^+ e^- \ra Z h h$}

The double differential cross section of the process $e^+ e^- \ra hh Z$,
Fig.~1b, is given by
\beq 
\frac{d\sigma (\ee \ra hhZ) }{dx_1 dx_2} = \frac{G_F^3 M_Z^6 }{384 
\sqrt{2} \pi^3 s} \, (a_e^2+v_e^2) \, \ \frac{ {\cal A} }{(1-\mu_Z)^2} 
\eeq
The couplings have been defined in the previous section.
$x_{1,2}=2E_{1,2}/ \sqrt{s}$ are the scaled energies of the Higgs
particles, $x_3=2-x_1-x_2$ is the scaled energy of the $Z$ boson;
$y_k=1-x_k$. The scaled masses squared are denoted by $\mu_i=M_i^2/s$.
In terms of these variables, the coefficient ${\cal A}$ in the cross
section may be written as: 
\beq
{\cal A} &=&  \left\{ \frac{a^2}{2} f_0 + 
\frac{ \sin^4(\beta-\alpha)}{4\mu_Z^2(y_1+\mu_h-\mu_Z)} \left[ \frac{f_1}
{y_1+\mu_h -\mu_Z} + \frac{f_2 }{y_2+\mu_h-\mu_Z} \right] + \frac{\cos^4
(\beta-\alpha)}{4\mu_Z^2(y_1+\mu_h-\mu_A)} \right. \non \\
&& \times \left[ \frac{f_3}{y_1+\mu_h-\mu_A}+\frac{f_4}{y_2+\mu_h-\mu_A} 
\right] + \frac{a}{\mu_Z} \left[ \frac{ \sin^2(\beta-\alpha) f_5}{y_1+\mu_h
-\mu_Z}+ \frac{\cos^2(\beta-\alpha) f_6} {y_1+\mu_h-\mu_A} \right]  \non \\
&&\left. + \frac{\sin^2 2(\beta-\alpha)}{8 \mu_Z^2(y_1+\mu_h-\mu_Z)} 
\left[ \frac{f_7}{y_1+\mu_h-\mu_Z} + \frac{f_8}{y_2+\mu_h-\mu_Z} \right] 
\right\} + \left\{ y_1 \leftrightarrow y_2 \right\}
\eeq
with 
\beq 
a = \frac{1}{2} \left[ 
\frac{\sin(\beta-\alpha) \lambda_{hhh} } {y_3+\mu_Z-\mu_h} +
\frac{\cos(\beta-\alpha) \lambda_{Hhh} } {y_3+\mu_Z-\mu_H} \right]
+ \frac{ \sin^2(\beta-\alpha)}{y_1+\mu_h-\mu_Z} + \frac{
\sin^2(\beta-\alpha) }{y_2+\mu_h-\mu_Z} + \frac{1}{2\mu_Z} 
\eeq
[omitting the small decay widths of the Higgs bosons]. Only the coefficient 
$a$ includes the Higgs self--couplings $\lambda_{Hhh}$ and $\lambda_{hhh}$. 
Introducing the notation $y_0=(y_1-y_2)/2$, the coefficients $f_i$ which do 
not involve any Higgs couplings, are defined by
\beq
f_0 & = & (y_1+y_2)^2-4\mu_Z(1-3 \mu_Z)   \\
f_1 & = & \left[ (1+y_1)^2 -4\mu_Z(y_1+\mu_h) \right] \left[
y_1^2+\mu_Z^2-2\mu_Z (y_1+2\mu_h) \right] \non \\
f_2 & = & \left[ 2\mu_Z(\mu_Z-2\mu_h +1)-(1+y_1)(1+y_2) \right] 
\left[ \mu_Z(\mu_Z-y_1-y_2-4\mu_h+2)- y_1 y_2 \right] \non \\
f_3 &=& \left[ y_0^2 + \mu_Z(1-y_1-y_2+\mu_Z-4\mu_h) \right] \left[
1+ y_1+y_2+y_0^2 + \mu_Z (\mu_Z -4\mu_h - 2y_1 ) \right] \non \\
f_4 & =& \left[ y_0^2 + \mu_Z (1-y_1-y_2+\mu_Z-4\mu_h) \right] 
\left[ y_0^2 -1+ \mu_Z(\mu_Z-y_1-y_2-4\mu_h+2) \right]  \non \\
f_5 &=& 2\mu_Z^3-4\mu_Z^2(y_1+2\mu_h)+\mu_Z \left[(1+y_1)(3y_1-y_2) +2 \right] 
-y_1^2(1+y_1+y_2) -y_1y_2 \non \\
f_6 &=& 2\mu_Z^3-\mu_Z^2(y_2+3y_1+8\mu_h-2)+2\mu_Zy_0 \left( 1+y_1+y_0 \right)+ 
2 y_1 y_0-y_0^2(y_1+y_2-2) \non \\
f_7 &=& \left[ \mu_Z(4\mu_h-\mu_Z-1+2y_1-y_0) -y_1y_0 \right] \left[
\mu_Z(4\mu_h-\mu_Z-1+3y_1) -(1+y_0)(1+y_1) \right] \non \\
f_8 &=& \left[ \mu_Z(4\mu_h-\mu_Z-1+2y_1-y_0) -y_1y_0 \right] \left[
\mu_Z(4\mu_h-\mu_Z-2+y_1) +(1-y_0)(1+y_1) \right] \non
\eeq
In the decoupling limit, the cross section is reduced to the $\SM$~cross section
for which 
\beq
{\cal A} = \frac{a^2}{2} f_0 +
\frac{1}{4\mu_Z^2(y_1+ \mu_h-\mu_Z)} \left[ \frac{ f_1}{y_1+\mu_h-\mu_Z}
+ \frac{f_2}{y_2+ \mu_h-\mu_Z} + 4a \mu_Z f_5 \right]  
+ \left\{ y_1 \leftrightarrow y_2 \right\} \non 
\eeq
with the $f_i$'s as given above, and 
\beq 
a = \frac{1}{2} \frac{ \lambda_{hhh} } {y_3+\mu_Z-\mu_h}
+ \frac{ 1} {y_1+\mu_h-\mu_Z} + \frac{1}{y_2+\mu_h-\mu_Z} 
+ \frac{1}{2\mu_Z} \non 
\eeq
The cross section $\sigma(\ee \ra hhZ)$ is shown for $\sqrt{s}=500$ GeV
at $\tb=1.5$ as a function of the Higgs mass $M_h$ in Fig.~4a. For small
masses, the cross section is built up almost exclusively by $H \ra hh$
decays [dashed curve], except close to the point where the
$\lambda_{Hhh}$ coupling accidentally vanishes (cf. Ref.\cite{R5}) and
for masses around $\sim 90$ GeV where additional contributions come from
the decay $A \ra hZ$ [this range of $M_h$ corresponds to $M_A$ values
where BR$(A \ra hZ)$ is large; c.f. Fig.3]. For intermediate masses, the
resonance contribution is reduced and, in particular above 90 GeV where
the decoupling limit is approached, the continuum $hh$ production
becomes dominant, falling finally down to the cross section for double
Higgs production in the Standard Model [dashed line]. After subtracting
the $H \ra hh$ decays [which of course is very difficult], the continuum
cross section is about $0.5$~fb, and is of the same order as the
\SM~cross section at $\sqrt{s}= 500$ GeV. Very high luminosity is
therefore needed to measure the trilinear $hhh$ coupling. At higher
energies, since the cross section for double Higgs--strahlung scales
like $1/s$, the rates are correspondingly smaller, c.f. Fig.4b. \s 

Prospects are similar for large $\tb$ values. The cascade decay $H \ra
hh$ is restricted to a small $M_h$ range of less than 70 GeV, with a 
production cross
section of $\sim 20$ fb at $\sqrt{s}=500$ GeV and $\sim 3$ fb at 1.5 TeV. The
continuum cross sections are of the order of $0.1$ fb at both energies, so that
very high luminosities will be needed to measure the continuum cross sections 
in this case if the background problems can be mastered at all. \s 

We have repeated the analysis for the continuum process $\ee \ra Ahh$
(cf. Fig.1c).
However, it turned out that the cross section is built up almost exclusively
by resonant $AH \ra Ahh$ final states, with a very small continuum
contribution, so that the measurement of the coupling $\lambda_{hAA}$ 
is extremely difficult in this process. 

\subsubsection*{3.2 $W_L W_L \ra hh $}

In the effective longitudinal $W$ approximation\footnote{For qualifying 
comments see footnote 3.}, the total cross 
section for the subprocess $W_L W_L  \ra hh$, Fig.~1d, is given by
\beq
\hat{\sigma}_{LL} &=& \frac{G_F^2 \hat{s}}{64 \pi} \frac{\beta_h}{ \beta_W} 
\left\{ (1+\beta_W^2)^2 
\left[\frac{\mu_Z \sin (\beta-\alpha) }{1-\mu_h} \lambda_{hhh} + 
\frac{\mu_Z \cos (\beta-\alpha) }{1-\mu_H} \lambda_{Hhh} +1 \right]^2 
\right. \\
&+& \frac{\beta_W^2} {\beta_W \beta_h} \left[\frac{\mu_Z \sin 
(\beta-\alpha) }{1-\mu_h} \lambda_{hhh} + \frac{\mu_Z \cos 
(\beta-\alpha) }{1-\mu_H} \lambda_{Hhh} +1 \right] [ 
\sin^2(\beta-\alpha) g_1  \non \\
&+& \cos^2 (\beta -\alpha) g_2]  \left. 
+\frac{1} {\beta_W^2 \beta_h^2} \left[ \sin^4(\beta-\alpha) g_3+
\cos^4(\beta-\alpha) g_4+ \sin^22(\beta-\alpha) g_5 \right] \right\} \non
\eeq
with 
\beq
g_1 &=& 2 [(\beta_W -x_W \beta_h)^2 +1-\beta_W^4] l_W
-4 \beta_h (2\beta_W -x_W \beta_h) \non \\
g_2 &=& 2 (x_C\beta_h -\beta_W)^2 l_C
+ 4 \beta_h (x_C\beta_h -2 \beta_W) \non \\
g_3 &=& \beta_h[\beta_hx_W (3 \beta_h^2 x_W^2+14\beta_W^2+2-2\beta_W^4)-
4\beta_W( 3 \beta_h^2 x_W^2 + \beta_W^2 +1 -\beta_W^4 )][l_W + x_Wy_W] \non \\
&&-  [ \beta_W^4 +(1-\beta_W^4)(1+2 \beta_W^2 -\beta_W^4)] [l_W/x_W-y_W]
- 2 \beta_h^2 y_W (2 \beta_W - \beta_h x_W)^2  \non \\
g_4 &=& \beta_h [ \beta_h x_C (3 \beta_h^2 x_C^2 +14 \beta_W^2) -4 
\beta_W (3 
\beta_h^2 x_C^2 + \beta_W^2) ] [l_C +x_C y_C] \non \\
&& -\beta_W^4 [l_C/x_C -y_C]-2y_C \beta_h^2 (2 \beta_W - \beta_h x_C)^2 \non \\
g_5&=& \frac{ \beta_h \beta_W l_W }{x_W^2-x_C^2} [2 x_W ( 2x_W^2 \beta_h 
\beta_W -x_C x_W^2 \beta_h^2 -x_C \beta_W^2 ) -2x_W^2 ( \beta_h^2 x_W^2 
+\beta_W^2 +1-\beta_W^4) \non \\
&&+ \frac{x_C}{\beta_W \beta_h } ( (\beta_h^2 x_W^2 + 
\beta_W^2)(1-\beta_W^4) 
+( \beta_h^2 x_W^2 +\beta_W^2 )^2)] - 4\beta_h^3 \beta_W (x_W+x_C) \non \\
&&+ \frac{ \beta_h \beta_W l_C }{x_C^2-x_W^2} 
[ 4x_C^3 \beta_h \beta_W -2x_C x_W( \beta_h^2 x_C^2 + \beta_W^2 
+1 -\beta_W^4) -2x_C^2 ( \beta_h^2 x_C^2 +\beta_W^2 ) \non \\
&&+ \frac{x_W}{\beta_W \beta_h } ( (\beta_W^2+\beta_h^2 x_C^2)(1-\beta_W^4) 
+( \beta_h^2 x_C^2 +\beta_W^2 )^2)]+2 \beta_H^2(x_Cx_W\beta_H^2+4\beta_W^2)
\eeq
The scaling variables are defined in the same way as before.  
$\hat{s}^{1/2}$ is the c.m. energy of the subprocess, $\beta_W=(1-4M_W^2/
\hat{s})^{1/2}$ and $\beta_h= (1-4M_h^2/\hat{s})^{1/2}$ are the velocities 
of the $W$ and $h$ bosons, and
\beq
x_W= (1-2 \mu_h)/(\beta_W \beta_h) \ \ ,  \ \ 
x_C= (1-2 \mu_h+2 \mu_{H^\pm} -2\mu_W )/(\beta_W \beta_h) 
\non \\
l_i =\log (x_{i}-1)/(x_i+1) \  \ , \ \ y_i= 2/(x_i^2-1) 
\hspace*{2cm}
\eeq
The value of the charged Higgs boson mass $M_{H^\pm}$ in the $H^\pm$ 
$t$--channel exchange diagram of Fig.1d is given by 
$M_{H^\pm}^2=M_A^2+M_W^2$. \s

In the decoupling limit, the cross section reduces again to the $\SM$ cross
section which in terms of $g_1$ and $g_2$, defined above, is given by: 
\beq
\hat{\sigma}_{LL} = \frac{G_F^2 \hat{s}}{64 \pi} \frac{\beta_h}{ \beta_W} 
\left\{ (1+\beta_W^2)^2 \left[\frac{\mu_Z \lambda_{hhh}}{1-\mu_h} +1\right]^2 
+\frac{1+\beta_W^2} {\beta_W \beta_h} \left[\frac{\mu_Z \lambda_{hhh}}
{1-h_1} +1 \right]g_1 +\frac{g_3} {\beta_W^2 \beta_h^2} \right\}
\eeq

After folding $\hat{\sigma}_{LL}$ with the longitudinal $W_LW_L$ luminosity
\cite{Wlumi}, 
one obtains the total
cross section $\sigma(\ee \ra \nu_e \bar{\nu}_e hh)$ shown in Fig.~4b
as a function of the light Higgs mass $M_h$ for $\tb=1.5$ at $\sqrt{s} =1.5$
TeV. It is significantly larger than for double Higgs--strahlung in the
continuum. Again, for very light Higgs masses, most of the events are $H \ra
hh$ decays [dashed line]. The continuum $hh$ production is of the same size as
pair production of \SM~Higgs bosons [dotted line] which, as anticipated,
is being approached near the upper limit of the $h$ mass in the decoupling
limit. The size of the continuum $hh$ fusion cross section renders this channel
more promising than double Higgs--strahlung for the measurement of the
trilinear $hhh$ coupling. \s 

For large $\tb$ values, strong destructive interference effects reduce the cross
section in the continuum to very small values, of order 10$^{-2}$ fb, before
the \SM~cross section is reached again in the decoupling limit. As before, the
$hh$ final state is almost exclusively built up by the resonance $H\ra hh$
decays. 

\subsection*{4. Summa}

It is convenient to summarize our results by presenting Fig.5, which
displays the areas of the $[M_A, \tb$] plane in which $\lambda_{Hhh}$
[solid lines, 135$^0$ hatching] and $\lambda_{hhh}$ [dashed lines, 
$45^0$ hatching] could eventually be accessible by experiment. 
The size of these areas is based on purely theoretical cuts so that
they are expected to shrink if background processes and detector
effects are taken into account. \s

$(i)$ In the case of $H \ra hh$, we require a lower limit of the cross
section $\sigma(H) \times$BR$(H \ra hh) >0.5$ fb and at the same time
for the decay branching ratio $0.1 < $BR$(H\ra hh)<0.9$, as discussed
earlier. Based on these definitions, $\lambda_{Hhh}$ may become
accessible in two disconnected regions denoted by I and II [$135^0$ 
hatched] in Fig.5. For low
$\tb$, the left boundary of Region I is set by LEP1 data. The gap
between Regions I and II is a result of the nearly vanishing
$\lambda_{Hhh}$ coupling in this strip. The right boundary of Region II
is due to the overwhelming $t\bar{t}$ decay mode for heavy $H$ masses,
as well as due to the small $H$ production cross section. For
moderate values of $\tb$, the left boundary of Region I is defined by
BR$(H\ra hh)>0.9$. In the area between Regions I and II, $H$ cannot
decay into two $h$ bosons, {\it i.e.} $M_H <2M_h$. For large $\tb \gsim
10$, BR$[H\ra hh (AA)]$ is either too large or too small, except in a
very small strip, $M_A \simeq 65$ GeV, towards the top of Region I. [Note that 
$h$ and $A$ are nearly  mass--degenerate in this area.]  \s

$(ii)$ The dashed line in Fig.5 describes the left boundary of the area
[$45^0$ hatched] in which $\lambda_{hhh}$ may become accessible; it is
defined by the requirement that the continuum $W_L W_L \ra hh$ cross
section, $\sigma_{\rm cont}$, is larger than $0.5$ fb. Note that the
resonant $H \ra hh$ events in Region II must be subtracted in order to
extract the $\lambda_{hhh}$ coupling. \s

In conclusion, we have derived the cross sections for the double
production of the lightest neutral Higgs boson in the \MSSM~at $\ee$
colliders: in the Higgs--strahlung process $\ee \ra Zhh$, [in the triple
Higgs production process $\ee \ra Ahh$], and in the $WW$ fusion
mechanism. These cross sections are large for resonant $H \ra hh$ decays
so that the measurement of the triple Higgs coupling $\lambda_{Hhh}$ is
expected to be fairly easy for $H\ra hh$ decays in the $M_H$ mass range
between 150 and 350 GeV for small $\tb$ values. The continuum processes
must be exploited to measure the triple Higgs coupling $\lambda_{hhh}$.
These continuum cross sections, which are of the same size as in the
\SM, are rather small so that high luminosities are needed for the
measurement of the triple Higgs coupling $\lambda_{hhh}$.

\vspace*{0.5cm}

\nn {\bf Acknowledgements:}  \s

\nn Discussions with G. Moultaka and technical
help by T. Plehn are gratefully acknowledged. A.D. thanks the Theory
Group for the warm hospitality  extended to him at DESY, and H.E.H.
acknowledges the partial support of the U.S. Department of Energy.

\vspace*{1mm}

\nn {\Large \bf Figure Captions}

\begin{itemize}

\item[{\bf Fig.~1:~}]
Main mechanisms for the double production of the light MSSM Higgs boson 
in $\ee$ collisions: a) $\ee \ra ZH$, $\ee \ra AH$ and $W W \ra H$ 
followed by $H \ra hh$; (b) $\ee \ra hhZ$, (c) $\ee \ra hhA$ and (d) $W W 
\ra hh$. 

\item[{\bf Fig.~2:~}]
The cross sections for the production of the heavy CP--even Higgs boson $H$ 
in $\ee$ collisions, $\ee \ra ZH/AH$ and $\ee \ra H\nu_e \bar{\nu}_e$, and for 
the background process $\ee \ra Ah$ [the dashed curve shows $\frac{1}{2}
\times \sigma(Ah)$ for clarity of the figures]. The c.m.~energies are 
chosen $\sqrt{s}=500$ GeV in (a), and 1.5 TeV in (b). 

\item[{\bf Fig.~3:~}]
The branching ratios of the main decays modes of the heavy CP--even neutral 
Higgs boson $H$ in (a), and of the pseudoscalar Higgs boson $A$ in (b).

\item[{\bf Fig.~4:~}]
The cross sections for $hh$ production in the continuum for $\tb=1.5$:
$\ee \ra hhZ$ at a c.m.~energy of $ \sqrt{s}=500$ GeV (a) and $W_L W_L \ra
hh$ at $\sqrt{s}=1.5$ TeV (b). 

\item[{\bf Fig.~5:~}]
The areas of the $[M_A, \tb$] plane in which the Higgs self--couplings
$\lambda_{Hhh}$  and $\lambda_{hhh}$ could eventually be accessible by 
experiment at $\sqrt{s}=1.5$ TeV [see text for further discussions].

\end{itemize}

%\end{document}

\newpage

\begin{figure}[htbp]
\centerline{\psfig{figure=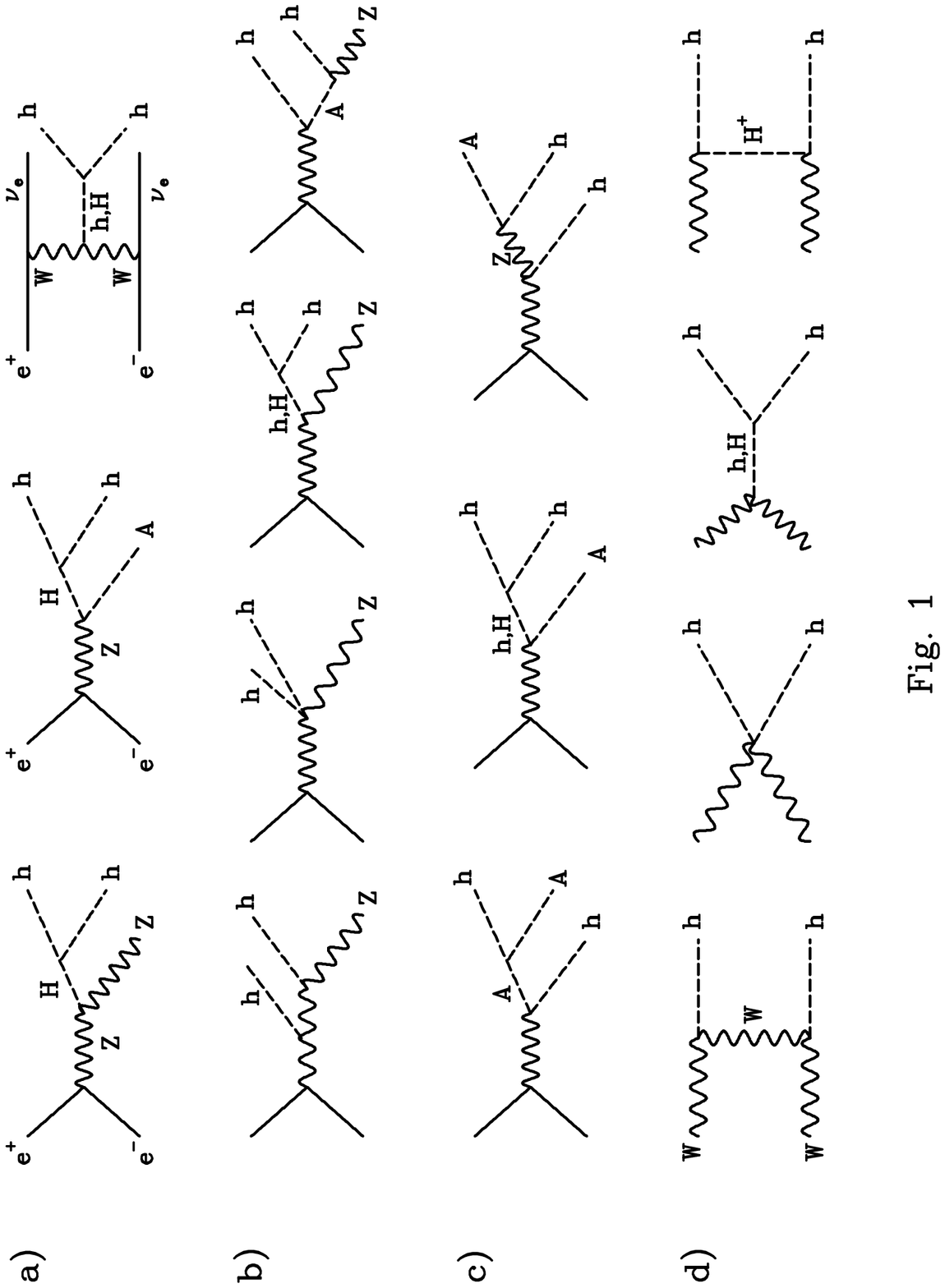,height=21.5cm,width=15cm}}
\vspace*{-1.2cm}
%\centerline{\bf Fig.~1}
\end{figure}

\newpage

\begin{figure}[htbp]
\centerline{\psfig{figure=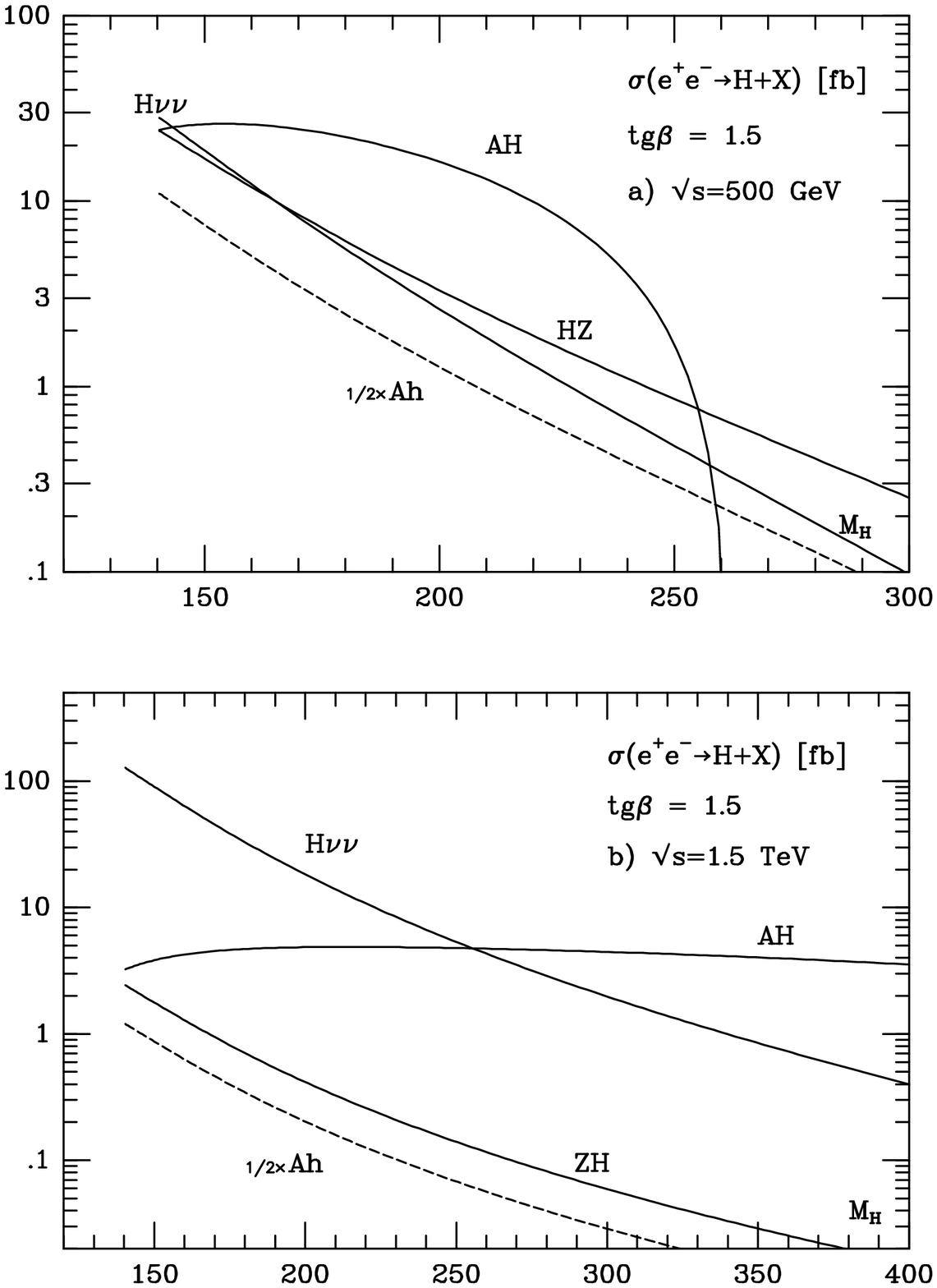,height=21.5cm,width=15cm}}
\vspace*{-1.2cm}
\centerline{\bf Fig.~2}
\end{figure}

\newpage

\begin{figure}[htbp]
\centerline{\psfig{figure=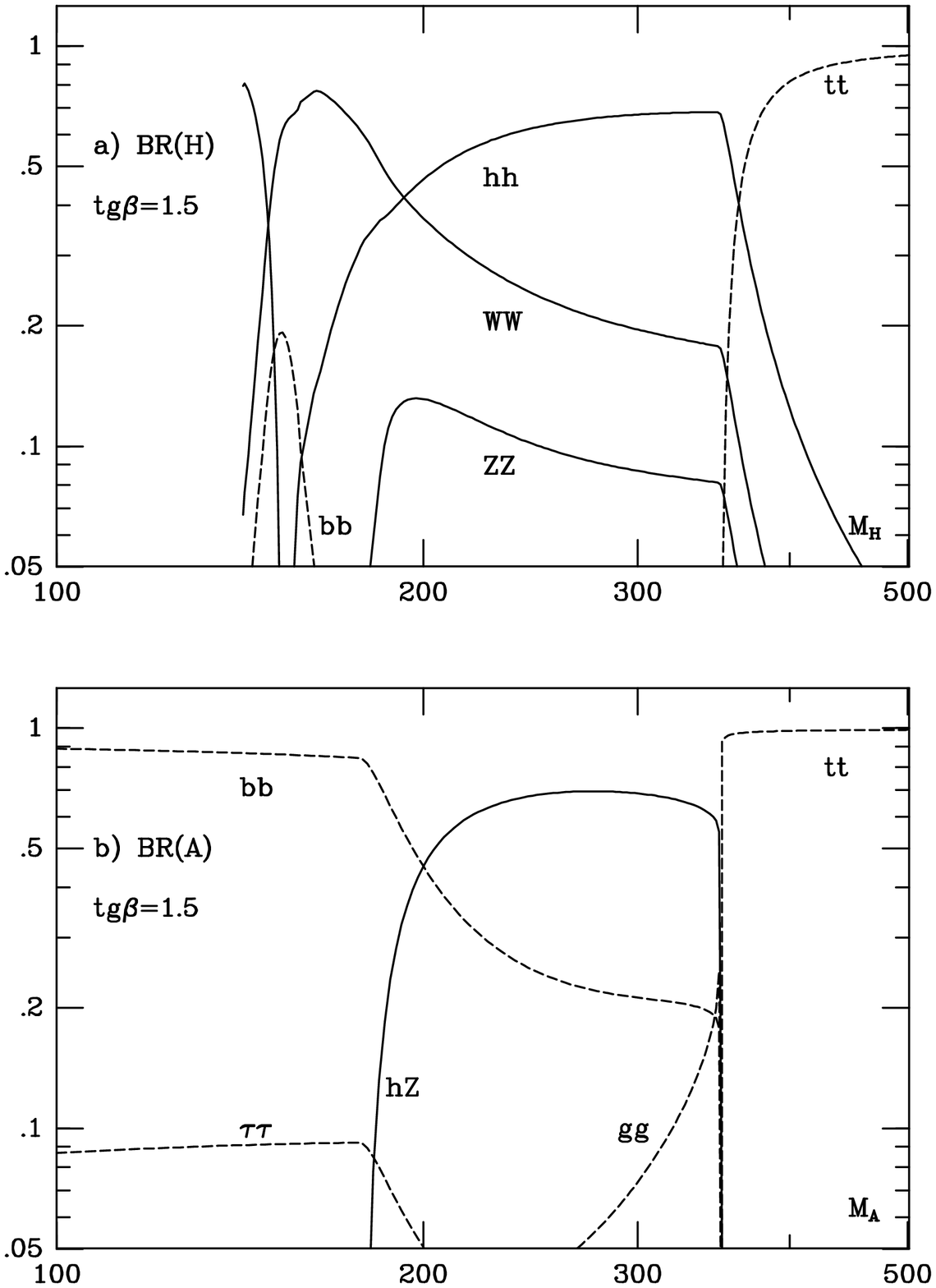,height=21.5cm,width=15cm}}
\vspace*{-1.2cm}
\centerline{\bf Fig.~3}
\end{figure}

\newpage

\begin{figure}[htbp]
\centerline{\psfig{figure=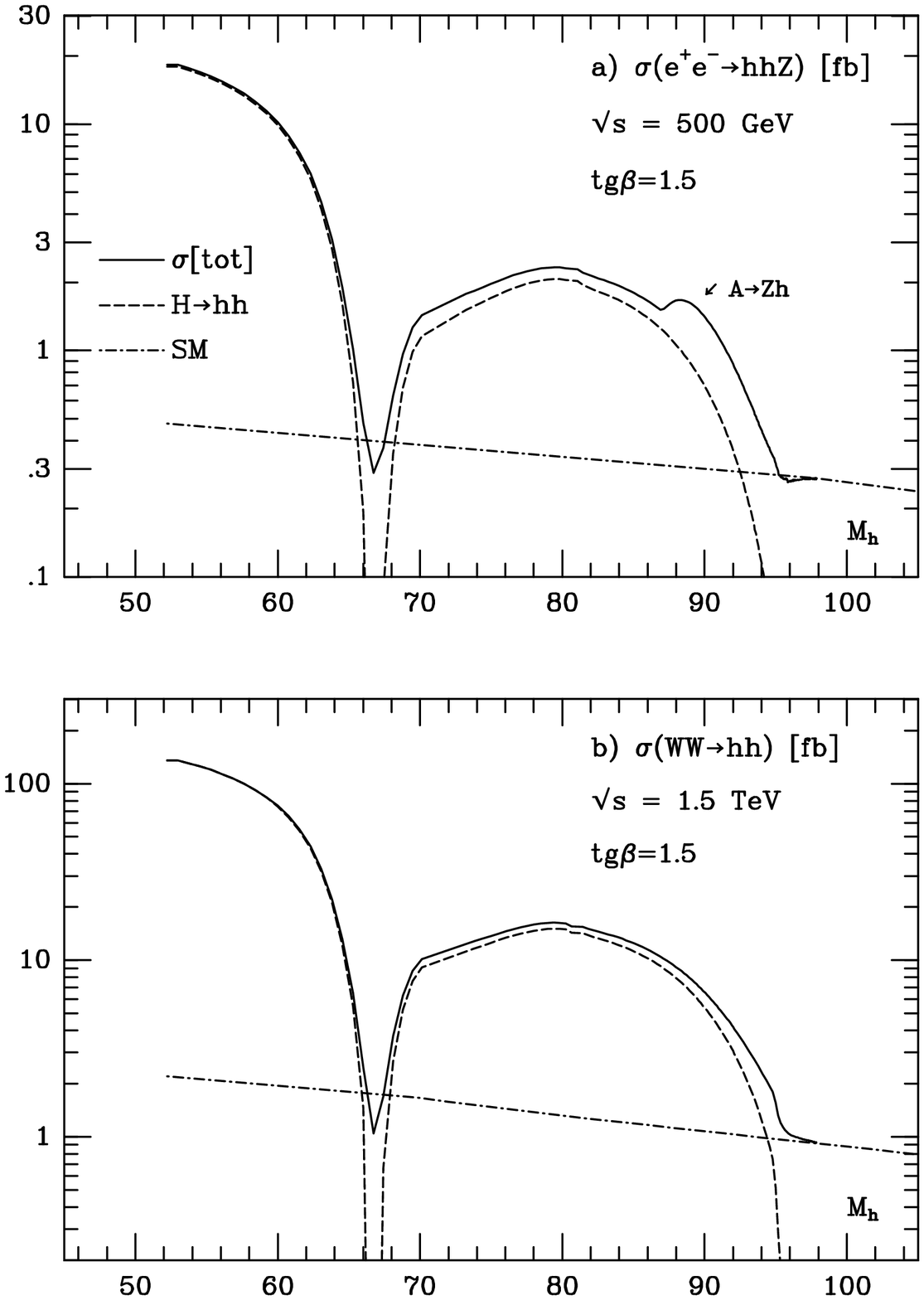,height=21.5cm,width=15cm}}
\vspace*{-1.2cm}
\centerline{\bf Fig.~4}
\end{figure}

\newpage

\begin{figure}[h]
\epsfig{file=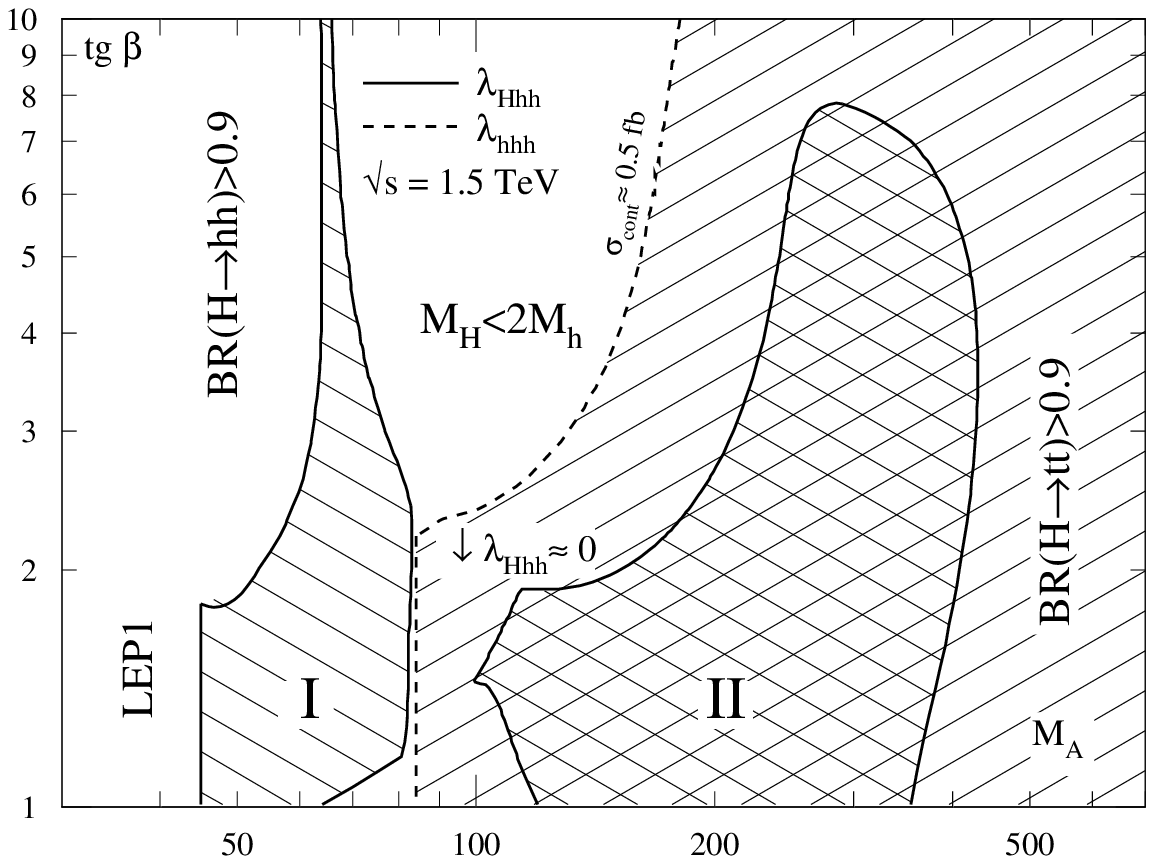,width=15cm}
%\caption[]{\bf Fig.~5}
%\label{\bf Fig.~5}
\vspace*{1.2cm}
\centerline{\bf Fig.~5}
\end{figure}

\end{document}